\documentclass[
  a4paper,
  fontsize=12pt,
  captions=tableheading,
  parskip=never,
  ]{scrartcl}
\setkomafont{caption}{\small}
\setkomafont{captionlabel}{\bfseries}
\addtokomafont{disposition}{\rmfamily\boldmath}
\addtokomafont{publishers}{\large}
\addtokomafont{subject}{\mdseries\large}

\setcapindent{1em}
\setcapdynwidth{0.9\textwidth}

\pdfoutput=1

\usepackage{mystyle}
\usepackage[mode=tex]{standalone}
\usepackage{subfiles}
\usepackage{import}

\graphicspath{ {./pictures_part1/} {./pictures_part2/} }
\DeclareGraphicsExtensions{.pdf,.png,.jpg}

\usetikzlibrary{arrows.meta,plotmarks}

\newcommand{\onlyinsubfile}[1]{#1}
\newcommand{\notinsubfile}[1]{}

\newcommand{\dEdx}{\ensuremath{\mathrm{d}E/\mathrm{d}x}}
\newcommand{\dedx}{\dEdx}
\newcommand{\mdEdx}{\ensuremath{\nicefrac{\mathrm{d}E}{\mathrm{d}x}}}
\newcommand{\mdedx}{\mdEdx}
\newcommand{\avg}[1]{\ensuremath{\left\langle #1\right\rangle}}
\newcommand{\abs}[1]{\ensuremath{\left\lvert #1\right\rvert}}

\newcommand{\rmsninety}{\ensuremath{{\text{RMS}_{90}}}}
\newcommand{\chem}[1]{\ensuremath{\mathrm{#1}}}

\newcommand{\ilcsoft}{iLCSoft}
\newcommand{\lctpc}{LCTPC}

\begin{document}

\hyphenation{
  am-pli-fi-ca-tion
  col-lab-o-ra-tion
  per-for-mance
  sat-u-rat-ed
  se-lect-ed
  spec-i-fied
}

\renewcommand{\onlyinsubfile}[1]{}
\renewcommand{\notinsubfile}[1]{#1}

\title{Studies on Particle Identification with \dedx{} for the ILD TPC}
\author{Uli Einhaus\thanks{Corresponding authors\protect\label{thanks}}
  \and Uwe Krämer
  \and Paul Malek\footref{thanks}}
\subject{Talk presented at the International Workshop on Future Linear Colliders (LCWS2018), Arlington, Texas, 22-26 October 2018. C18-10-22.}
\publishers{Deutsches Elektronen\-/Synchrotron, DESY}
\date{}

\maketitle

\begin{abstract}
  \noindent
  For the International Large Detector (ILD) at the planned International Linear Collider (ILC) a Time Projection Chamber (TPC) is foreseen as the main tracking detector.
  To achieve the required point resolution, Micro Pattern Gaseous Detectors (MPGD) will be used in the amplification stage.
  A readout module using a stack of three Gas Electron Multipliers (GEM) for gas amplification was developed at DESY.
  In a test campaign at the DESY II Test Beam Facility the performance of three of these modules was investigated.
  In this contribution results were presented on the particle
  identification capabilities of the system using the specific energy loss (\dedx).
  The results from the prototype were used to extrapolate to the performance of the full ILD TPC, where a \dedx{} resolution of better than \SI{5}{\percent} could be achieved.
  In addition, simulation studies were performed to optimise the readout pad size for improved \dedx{} separation power.
  These studies also investigated the possibility to measure the deposited energy by counting the number of ionisation clusters (cluster counting).
  For small enough pads this approach was found to give similar or better performance compared to the traditional method of measuring the deposited charge.
\end{abstract}

\clearpage
% \flushbottom

\tableofcontents

\section{Project Background}
\label{sec:background}

This R\&D project is part of the design effort for the International Large Detector (ILD,~\cite{ild_loi}) at the planned International Linear Collider (ILC,~\cite{ilc_tdr}).
The FLC TPC group at DESY is a member of the \lctpc{} collaboration~\cite{lctpc_www}, which is driving the efforts of developing a Time Projection Chamber (TPC) for the ILD.
Different readout and amplification technologies based on micro pattern gaseous detectors (MPGD) are currently studied within the collaboration.
FLC TPC is developing a readout module with a gas amplification stage composed of Gas Electron Multipliers (GEM,~\cite{sauli1997}) mounted on thin ceramic grids.

The feasibility of using GEM mounted on thin ceramic grids in a TPC amplification stage was first demonstrated with \SI{10x10}{\centi\meter} GEM~\cite{steder2013}.
In the next step, measurements were performed at the DESY II test beam with a series of larger readout modules.% (see \cref{sec:tb:module}).
With this data it was shown that the design goals for the ILD TPC regarding spacial resolution could be reached~\cite{lctpc2016,mueller2016}.
A new generation of three modules was build for the beam tests described in \cref{sec:testbeam}, using improved tooling described in~\cite{malek2017}.

\section{Determining \dedx{} Resolution from Beam Tests}
\label{sec:testbeam}

Beam tests were performed at the DESY II Test Beam Facility with a large TPC prototype inside a \SI{1}{\tesla} solenoid magnet (\cref{sec:tb:testbeam}).
Three readout modules with a GEM gas amplification stage were used (\cref{sec:tb:module}) in the measurements for the determination of the \dedx{} resolution (\cref{sec:tb:measurement}).

\subsection{The DESY GEM Module}
\label{sec:tb:module}

\begin{figure}[hbp]
  \begin{subfigure}[t]{0.5\textwidth}
    \centering
    \includegraphics[clip=true,
                     trim=190mm 260mm 100mm 90mm,
                     width=\textwidth,
                     height=0.33\textheight,
                     keepaspectratio=true,
                     ] {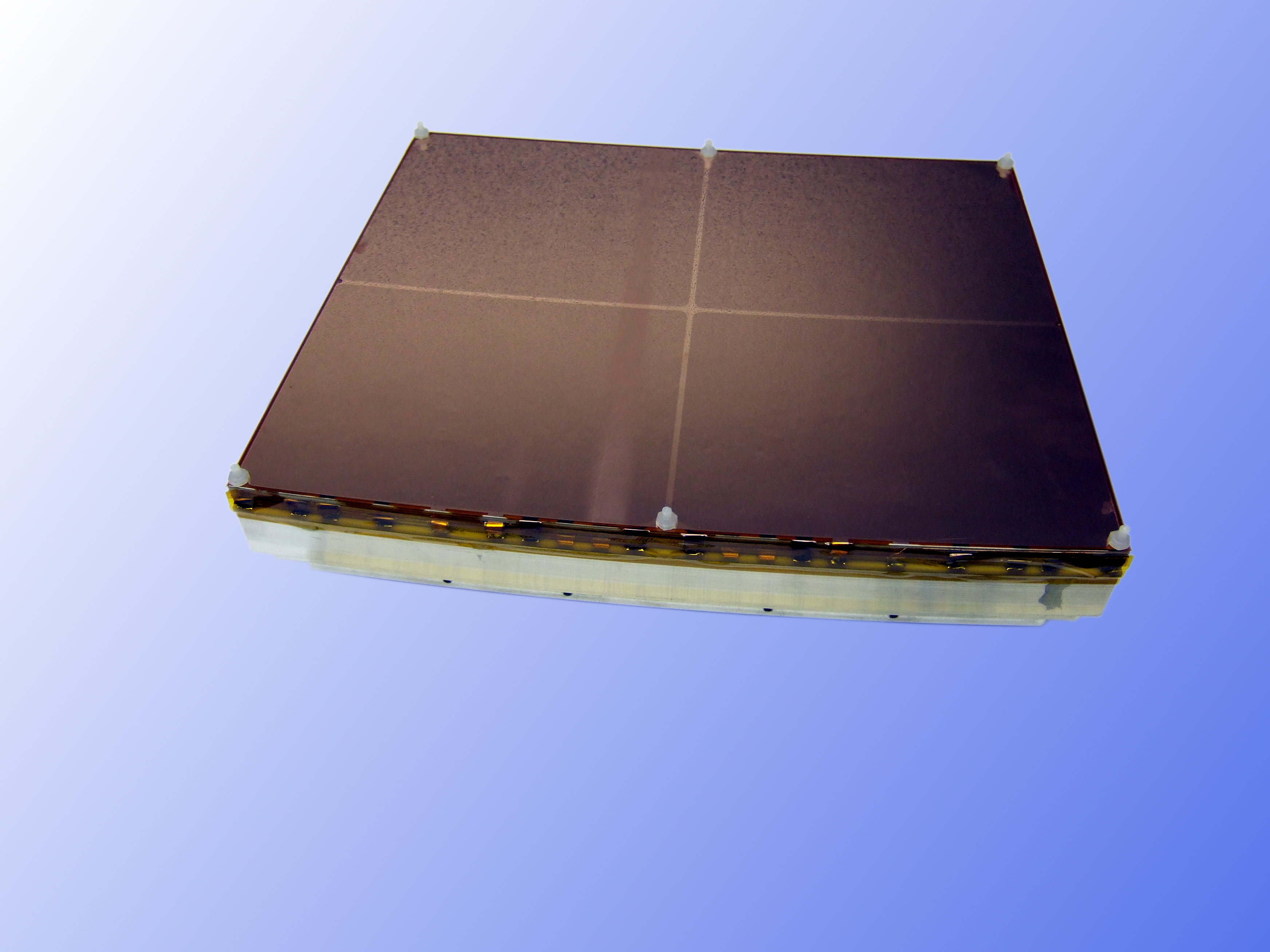}
    \caption{A fully assembled GEM module.
      \label{fig:tb:module}}
  \end{subfigure}%
  \begin{subfigure}[t]{0.5\textwidth}
    \centering
    \includestandalone[width=\textwidth,
                       height=0.33\textheight,
                       keepaspectratio=true
                       ] {module_explosion}
    \caption{Explosion view of the module.\footnotemark
      \label{fig:tb:module_explosion}}
  \end{subfigure}
  \caption{The DESY GEM module and its major components.
    \label{fig:tb:module_overview}}
\end{figure}

In \cref{fig:tb:module_overview} the GEM module and its major components are shown.
The footprint of a module is a sector of an annulus with an inner radius of \SI{1400}{\milli\meter} and an outer radius of \SI{1600}{\milli\meter} encompassing about \ang{8.4}, similar to the geometry of the readout modules intended for the ILD TPC.
The mechanical base of the module is an aluminium back frame, which is used to mount it in the TPC end plate.
A printed circuit board carrying the segmented readout anode is glued onto the back frame.
The \num{4832} readout pads are arranged in \num{28} concentric rows with a pitch of \SI{1.2625}{\milli\meter} along the rows and \SI{5.85}{\milli\meter} row pitch.
On top of the pad board, the amplification structure is build up out of three GEMs supported by \SI{1}{\milli\meter} high alumina\-/ceramic ($\mathrm{Al_2O_3}$) frames.
Additional frames serve as spacers between the GEM foils to define the transfer and induction gaps of \SIlist{2;3}{\milli\meter}, respectively, as shown in \cref{fig:tb:module_explosion}.
Since the frame bars are only \SI{1.4}{\milli\meter} wide and the space needed for the high voltage connections of the GEMs was kept at a minimum, the module has a sensitive area of over \SI{90}{\percent}.
The readout electronics are based on the ALICE TPC ReadOut (ALTRO) analog\-/to\-/digital converter (ADC) chip and the PCA16 pre\-/amplifier~\cite{altro2008,eudetpca16}.
The ALTRO chip provides digitisation with \SI{10}{\bit} resolution at sampling frequencies of \SIlistor{5;10;20}{\mega\hertz}.
At \SI{40}{\MHz} the effective resolution is reduced to about \SI{8.5}{\bit}.
For the beam tests it was run at \SI{20}{\MHz} to make use of the full resolution.

\footnotetext{\label{else_perm}With permission from~\cite{lctpc2016}, Elsevier, 2016}

\subsection{The LCTPC Prototype Setup}
\label{sec:tb:testbeam}

For the development of the different technologies the LCTPC collaboration build a common setup at the beam line T24/1 at the DESY II Test Beam Facility.
The beam line delivers electrons with energies up to \SI{6}{\giga\electronvolt} at rates of about \SI{100}{\hertz} and up to several \si{\kilo\hertz} for lower energies~\cite{desy_testbeam2018}.
Four scintillator fingers are placed into the beam to provide a coincidence signal as a trigger for the main setup.

\begin{figure}[hbp]
  \begin{subfigure}{0.5\textwidth}
    \centering
    \includegraphics[clip=true,trim=80mm 225mm 65mm 200mm,width=\textwidth,height=0.33\textheight,keepaspectratio=true]{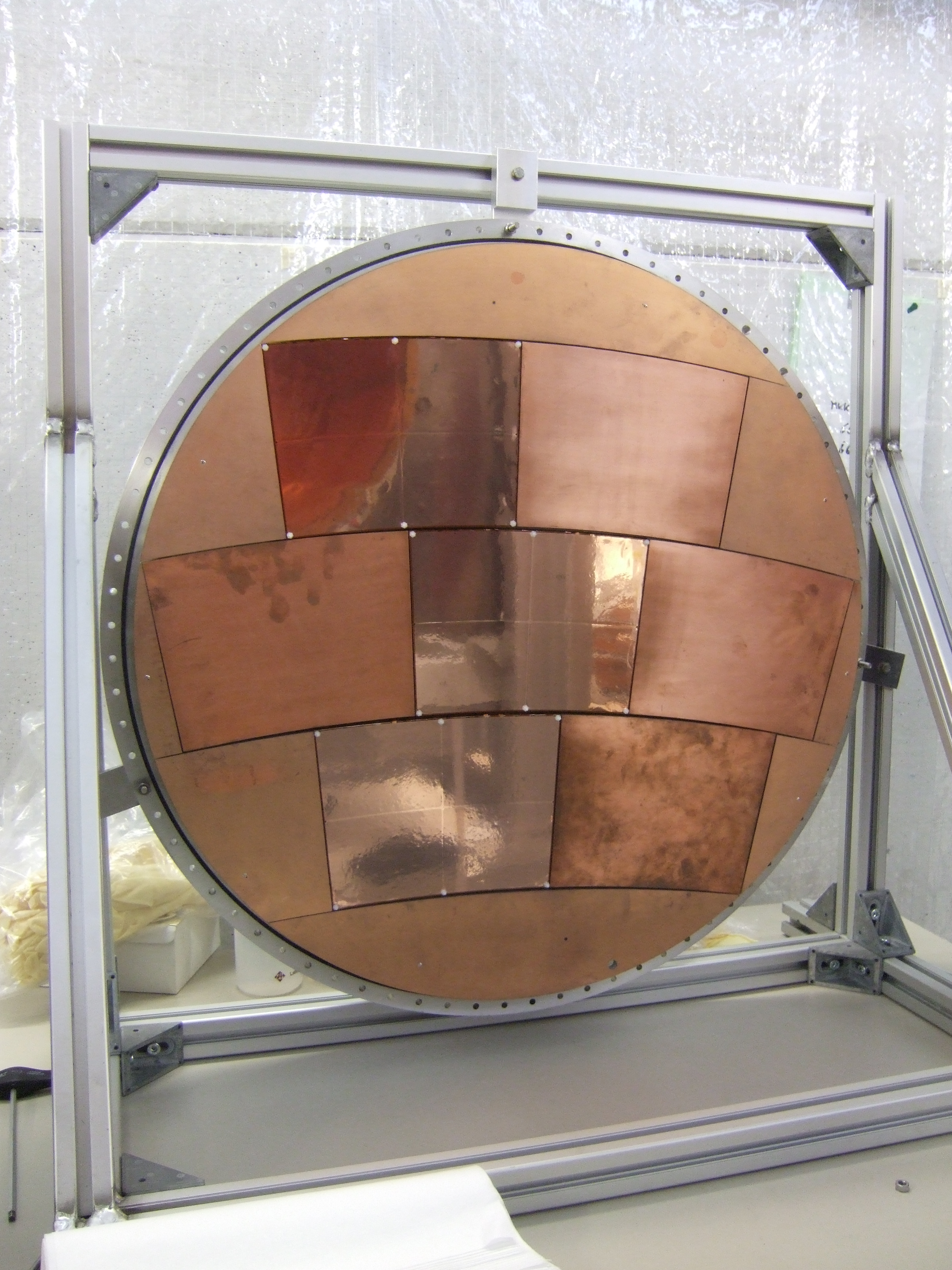}
    \caption{LPTPC anode end\-/plate.\footnotemark
      \label{fig:endplate}}
  \end{subfigure}%
  \begin{subfigure}{0.5\textwidth}
    \centering
    \includegraphics[clip=true,trim=0mm 20mm 0mm 20mm,width=\textwidth,height=0.33\textheight,keepaspectratio=true]{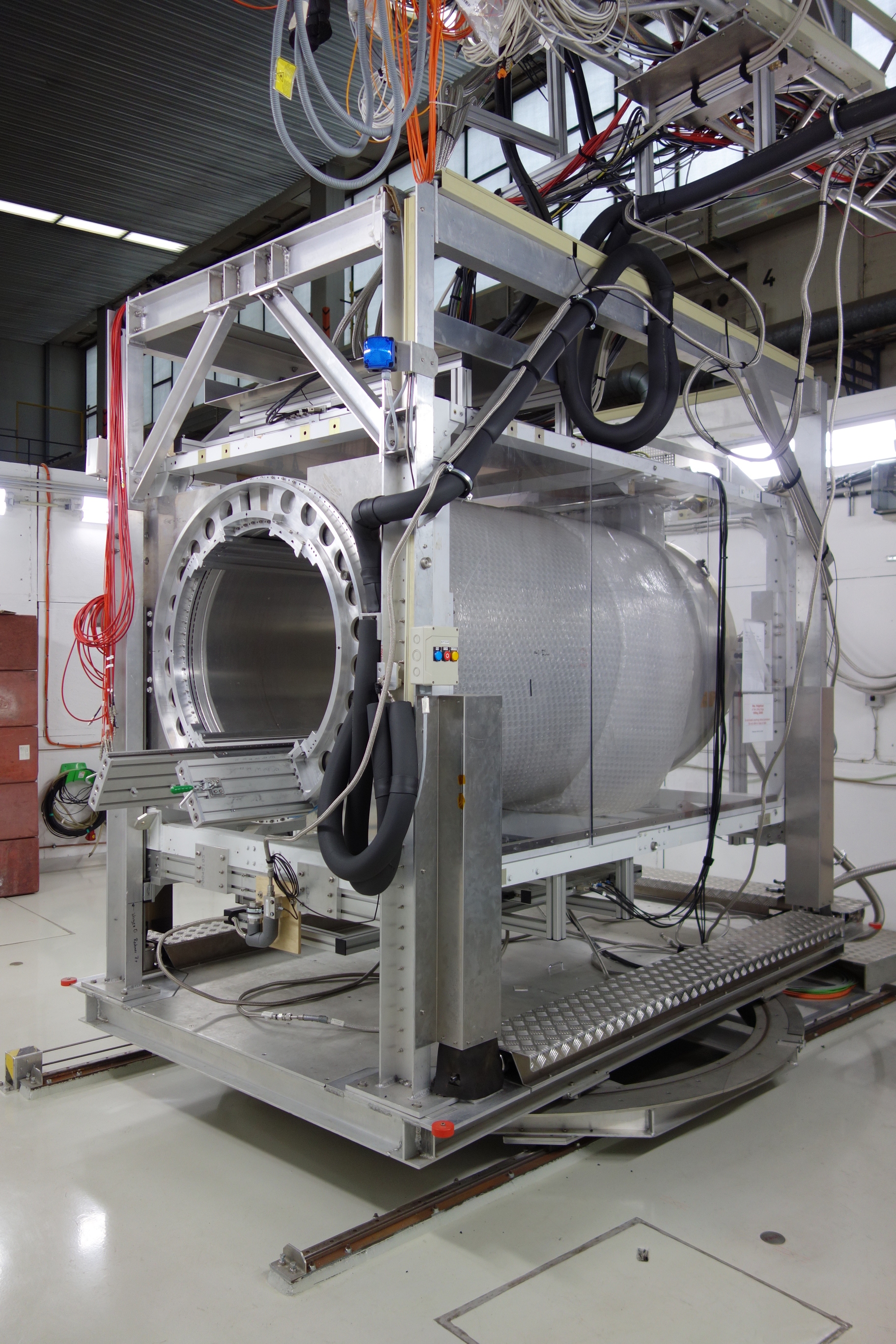}
    \caption{PCMag in area T24/1 at DESY II.
      \label{fig:magnet}}
  \end{subfigure}
  \caption{The \lctpc{} test beam setup.
    (\subref*{fig:endplate})~The inside of the anode end plate of the LPTPC with its seven module slots.
    The three mounted modules are distinguishable by their shinier surface.
    (\subref*{fig:magnet})~The PCMag on its movable stage.}
\end{figure}

The first part of the setup is a large TPC prototype (LPTPC) with about \SI{570}{\milli\meter} maximum drift distance and a diameter of \SI{75}{\centi\meter}.
The anode end plate shown in \cref{fig:endplate} provides slots for seven readout modules arranged in three rows.
The measurable track length perpendicular to these rows is about \SI{500}{\milli\meter}.
Unused slots can be closed with dummy modules for an uniform termination of the drift field.
The field\-/cage wall constitutes a little more than \SI{1}{\percent} of a radiation length $X_0$~\cite{schade2010lptpc}.

A superconducting solenoid magnet generating a \SI{1}{\tesla} magnetic field, the so\=/called Persistent Current Magnet (PCMag), is installed in the experimental area of the beam line.
As shown in \cref{fig:magnet}, it is mounted on a movable stage, which can be translated vertically and horizontally perpendicular to the beam axis, as well as rotated around the vertical axis to change the angle of the beam inside the TPC.
The usable inner diameter of the magnet measures \SI{85}{\centi\meter}.
Due to a lightweight construction without a field return yoke, the magnet walls make up only about \SI{20}{\percent} of a radiation length~\cite{pcmag1994}.
This is especially important as the electron beam has to pass through the magnet wall to reach the TPC inside and any material in front of the TPC deteriorates the beam quality.

\footnotetext{With permission from~\cite{lctpc2016}, Elsevier, 2016}

A gas system, described in \cite{schaefer2006}, provides a constant flow rate of the chosen gas, which in this tests is the gas mixture used in the TPCs in the near detector of the T2K experiment~\cite{t2k2011}.
The T2K gas mixture with the three components argon, tetrafluoromethane (\chem{CF_4}) and 2\=/methylpropane (\emph{isobutane}; \chem{HC(CH_3)_3}) in a ratio of $95:3:2$, was chosen as a candidate for the ILD TPC because of its high drift velocity and low diffusion allowing for good spacial resolution at long drift distances, as well as its high defocussing in the amplification region for improved charge sharing.

The beam test was performed with a nominal beam energy of \SI{5}{\giga\electronvolt}.
The drift field was chosen as \SI{240}{\volt\per\centi\meter}, close to the maximum of the drift velocity in T2K gas.

\subsection{Reconstruction \& Analysis}
\label{sec:tb:measurement}

For the reconstruction and analysis of the beam test data packages from the the \ilcsoft{} environment are used~\cite{ilcsoft2007}.
The raw data from the readout electronics are converted into the LCIO format~\cite{lcio2012}.
Then the reconstruction and analysis of the data are performed with Marlin \cite{marlin2006}, using the processors and scripts in the MarlinTPC package~\cite{marlintpc2007}.

The raw data consist of a stream of digitised charge values for each channel.
In the reconstruction each stream is searched for an excess of charge consistent with the arrival of a signal, so\=/called pulses.
Signal like pulses are required to exceed a minimum threshold for a limited amount of samples.
The pulses contain the full time information, the only spacial information is the pad position.
Hits are created from neighbouring pulses by searching for local maxima inside one pad row and combining pulses around these.
An accurate position parallel to the row is calculated as the charge weighted mean position of the contributing pulses.
Since most tracks are expected to cross the rows mostly perpendicular, the position within the row pitch is assumed as the centre of the row.
The arrival time of the maximum charge pulse and the known drift velocity are used to determine the position in the drift direction.
The charge of the hit is set to the sum of the charge of the corresponding pulses.
If the hit includes a pulse, in which a second rise was detected, it is flagged as a multi\-/hit candidate.
Finally, a track finder based on a Hough transformation algorithm is applied on the hit level~\cite{kleinwort_hough}, and all found tracks are subsequently fitted using the General Broken Lines method~\cite{kleinwort_gbl}.

\subsubsection{Determination of \dedx{} Resolution}
\label{sec:tb:analysis}

To avoid contamination of the charge values from overlapping tracks, the \dedx{} analysis is only performed on events with a single reconstructed track.
As some of the beam electrons lose a lot of energy in the magnet wall and the energy loss curve for electrons drops significantly below energies of about \SI{1}{\giga\electronvolt}, an equivalent cut on the track curvature is applied.

\begin{figure}[tb]
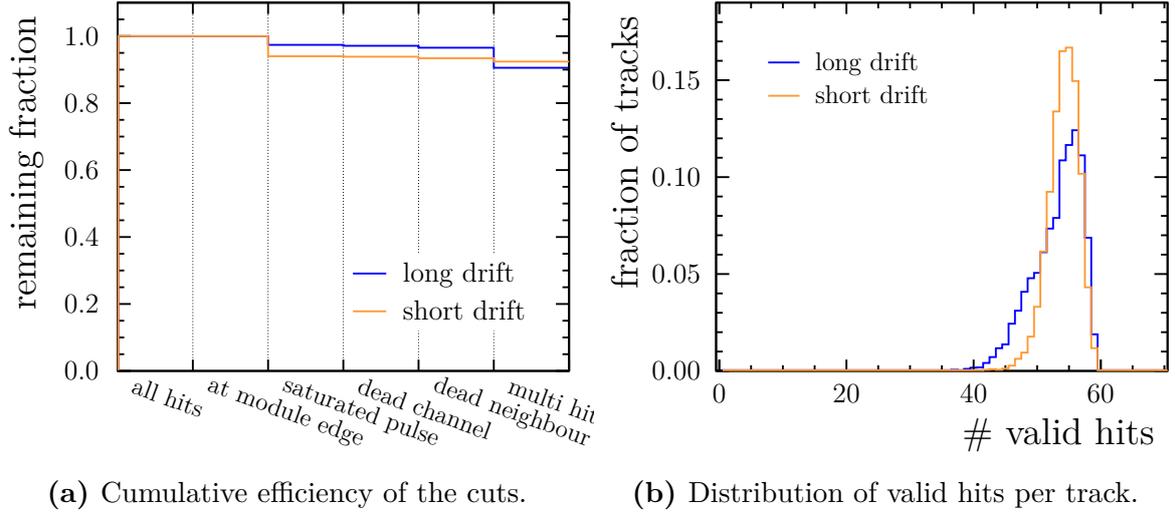

  \begin{subfigure}[t]{0.5\textwidth}
%     \centering
    \includestandalone[%
      width=\textwidth,
      height=0.33\textheight,
      keepaspectratio=true,
      ] {hit_cut_flow}
    \caption{Cumulative efficiency of the cuts.
      \label{fig:hit_cut_flow}}
  \end{subfigure}%
  \begin{subfigure}[t]{0.5\textwidth}
    \includestandalone[%
      width=\textwidth,
      height=0.33\textheight,
      keepaspectratio=true,
      ] {num_est}
    \caption{Distribution of valid hits per track.
      \label{fig:num_hits}}
  \end{subfigure}
  \caption{Effect of the cuts on hit quality applied for the determination of the mean \dedx{} of a track.
    Short and long drift are \SIlist{20;500}{\milli\meter}, respectively.
    (\subref*{fig:hit_cut_flow})~The cumulative efficiency of the cuts.
    The \emph{at module edge} cut does not include hits on the first and last row of a module, as those are treated separately.
    (\subref*{fig:num_hits})~The resulting distribution of the number of valid hits on each track.
    \label{fig:hit_cuts}}
\end{figure}

The average energy loss on a track is calculated from the charge of all hits, that were linked to the track by the track finder.
To reduce biases from the measured charge values the hits need to fulfil the following quality criteria:
\begin{itemize}
  \item Not be at the module edge, as charge could not be fully recorded.
  \item Contain no pulses that saturated the range of the digitiser.
  \item Contain no dead channels or be right next to one.
  \item No multi\-/hit candidate as determined by the pulse finder based on the pulse shape.
\end{itemize}
\Cref{fig:hit_cuts} shows the cumulative efficiency of these cuts for two different drift distances and the resulting distribution of the number of valid hits on each track.
As detailed in \cref{sec:tb:estimators}, the rejection of hits with saturated pulses introduces a bias depending on the drift distance.
As the charge cloud is spread over more pads due to diffusion, the average charge per pulse is reduced.
Since the average hit charge stays the same, this cut removes more hits with lower charge at short drift distances than at long drift distances.
To minimise the influence of this cut, the results shown here were taken a large drift distance, were the fraction of saturated pulses is lowest.
Regarding the multi\-/hit cut, at the time of performing this analysis the hit finder algorithm did not identify multi\-/hit candidates in the readout plane along a pad row. Only tagging in drift direction is done by the pulse finder as mentioned in the introduction of \cref{sec:tb:measurement}.

\begin{figure}[htp]
  \begin{subfigure}{0.5\textwidth}
%     \centering
    \includestandalone[%
                       width=\textwidth,
                       height=0.33\textheight,
                       keepaspectratio=true,
                       ]{dEdx_dist}
      \caption{Distribution of \dedx{} of hits.
      \label{fig:hitcharge}}
  \end{subfigure}%
  \begin{subfigure}{0.5\textwidth}
%     \centering
    \includestandalone[%
                       width=\textwidth,
                       height=0.33\textheight,
                       keepaspectratio=true,
                       ]{dEdx_dist_trans}
      \caption{\dedx{} of hits transformed as \cref{eqn:tb:dedx_trans}.
      \label{fig:hitcharge_trans}}
  \end{subfigure}
  \caption{The \dedx{} distribution of all hits after cuts.
    (\subref*{fig:hitcharge})~The untreated distribution shows the long Landau\-/like tail to high values.
    (\subref*{fig:hitcharge_trans})~Transforming the \dedx{} values according to \cref{eqn:tb:dedx_trans} compresses the tail and results in a more symmetric distribution.
    \label{fig:hitcharge_comp}}
\end{figure}

\begin{figure}[tp]
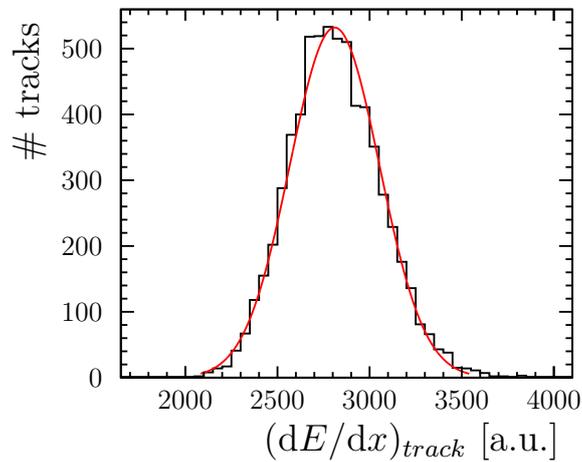

  \centering
  \includestandalone[%
    width=0.5\textwidth,
    height=0.33\textheight,
    keepaspectratio=true,
    ]{dEdx_mean}
  \caption{Distribution of the average \dedx{} of tracks.
    A Gaussian fit was applied in a $3\sigma$-range around the mean.
    The $\chi^2/ndf = \num{52.36/26}$ of the fit reflects the slight asymmetry visible in the distribution.
  \label{fig:track_dedx}}
\end{figure}

The energy loss per track length \dedx{} for each hit is calculated by dividing the hit charge $Q_{hit}$ by the length of the reconstructed track segment $\Delta s$ above the pad row corresponding to the hit.
\begin{equation}
  \label{eqn:dedx_hit}
  \left(\mdedx\right)_{hit} = \frac{Q_{hit}}{\Delta s}.
\end{equation}
The \dedx{} spectrum of all hits remaining after cuts is depicted in \cref{fig:hitcharge}.
As expected it shows a Landau\-/like distribution with a long tail towards high charge values, which is suppressed by the dynamic range of the readout electronics.
To suppress the influence of these hits, adapting the method used in \cite{kleinwort2008} each hit charge is transformed according to
\begin{equation}
  \label{eqn:tb:dedx_trans}
  \left(\mdedx\right)_{trans} = \frac{1}{\sqrt{\left(\mdedx\right)_{hit}}},
\end{equation}
which results in a narrower and more symmetric distribution, shown in \cref{fig:hitcharge_trans}.
To calculate the average energy loss of the track the mean of the transformed \dedx{} values of the connected hits is taken and the inverse transformation is applied:
\begin{equation}
  \label{eqn:tb:inverse_trans}
  \left(\mdedx\right)_{track} = \left(\frac{\sum_{hits} \left(\mdedx\right)_{trans}}{N_{hits}}\right)^{-2}.
\end{equation}

The distribution of the track mean \dedx{} values, shown in \cref{fig:track_dedx}, is almost Gaussian except for a small residual tail towards higher values, most likely arising from untagged multi hits.
For this reason the \dedx{} resolution is not calculated from the fit parameters, but from the RMS of the smallest interval containing \SI{90}{\percent} of the entries (\rmsninety).
The relative resolution in this sample after applying the calibrations from \cref{sec:tb:calibration} was thereby determined to be \SI{8.7+-0.1}{\percent}.

\subsubsection{Charge Calibration}
\label{sec:tb:calibration}

\newcommand{\slopecc}{\ensuremath{s_{cc}}}
\newcommand{\offsetcc}{\ensuremath{o_{cc}}}
\newcommand{\sloperow}{\ensuremath{f_{row}}}

In an attempt to imitate the calibrations done for the charge measurement in the final detector, two correction methods were devised.
Since the electronics do not provide the possibility for a self\-/calibration of the readout channels, a similar result could be achieved by pulsing the common electrode of the GEM directly above the readout pads.
By varying the amplitude of the pulses the charge induced in the pads changes accordingly, and a calibration factor \slopecc{} and offset \offsetcc{} for each channel was gained.
The pulse charge is then corrected as $q_{cor} = (q_{meas} / \slopecc) - \offsetcc$.
Due to the difference in permittivity this calibration is not valid for channels close to the ceramic frames.
Therefore, all rows next to the frame were ignored in all analyses.

The second correction targets local inhomogeneities of the gas amplification.
Since the electron beam is not moved during data taking, most tracks pass over each row in about the same spot.
Therefore, a correction factor can be calculated for each row instead of each pad.
To avoid biases from the local track angle, the \dedx{} value of the hits is used instead of the raw charge.
The correction factor for a row \sloperow{} is then calculated from a subset of the data as the mean on this row $\avg{\mdedx}_{row}$ divided by the average of all rows:
\begin{equation}
  \label{eqn:tb:calibration_row}
  \sloperow = \frac{\avg{\mdedx}_{row}}{\sum_{rows}\avg{\mdedx}_{row}/N_{rows}}.
\end{equation}
The hit charge in the rest of the data is accordingly corrected as $Q_{hit}^{cor} = Q_{hit} / \sloperow$.

\begin{figure}[tbp]
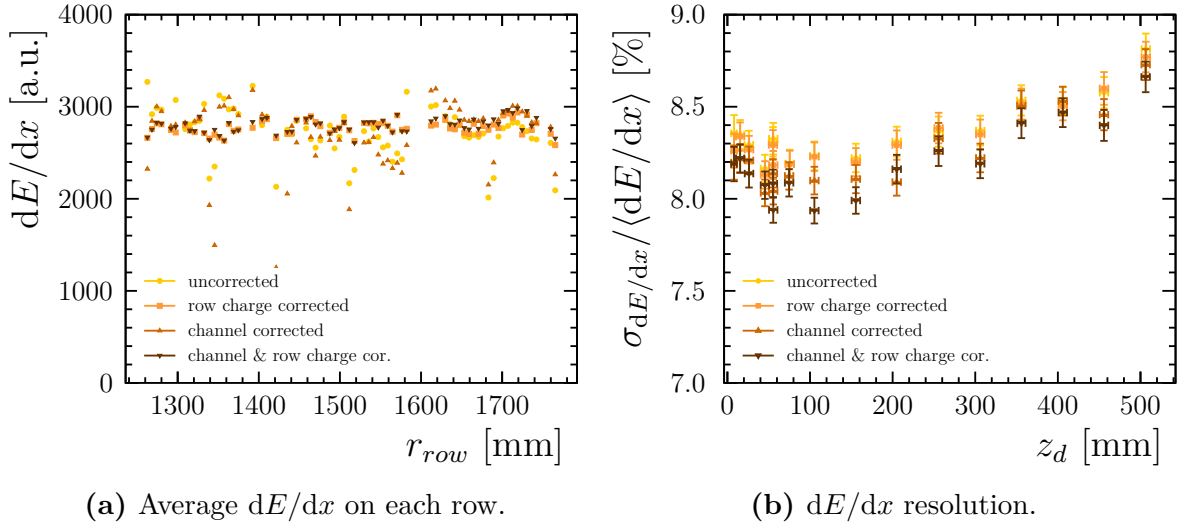

  \begin{subfigure}{0.5\textwidth}
%     \centering
    \includestandalone[%
                       width=\textwidth,
                       height=0.33\textheight,
                       keepaspectratio=true,
                       ]{cal_dEdx_row}
      \caption{Average \dedx{} on each row.
      \label{fig:calibration_row}}
  \end{subfigure}%
  \begin{subfigure}{0.5\textwidth}
%     \centering
    \includestandalone[%
                       width=\textwidth,
                       height=0.33\textheight,
                       keepaspectratio=true,
                       ]{cal_dEdx_res}
      \caption{\dedx{} resolution.
      \label{fig:calibration_res}}
  \end{subfigure}
  \caption{(\subref*{fig:calibration_row})~The average hit \dedx{} on each row with different applied corrections.
    (\subref*{fig:calibration_res})~The resulting resolution at each measured drift distance.
    The increase towards higher drift distances is caused by the changing fractions of saturated pulses and multi\-/hits, as well as electron attachment due to oxygen contamination.
    \label{fig:calibration}}
\end{figure}

\Cref{fig:calibration_row} shows the impact of both corrections on the spread of the mean charge between individual rows.
It is clearly visible, that the correction of the average row charge leads to a much more uniform response.
Especially the rows at the module edges and next to the ceramic frames, which show much lower charge, are equalised.
The influence on the \dedx{} resolution is only minor, as could be expected, since the total gain inhomogeneities are still much smaller than the fluctuations of the primary ionisation.
Still, a slight improvement can be seen in \cref{fig:calibration_res} when both corrections are applied.
To make the results comparable the average offset from the channel correction was applied to the uncorrected and the row charge corrected data.
Otherwise, the resulting shift of the \dedx{} mean value would lead to a seemingly worse performance of the channel correction compared to uncorrected data.

\subsubsection{Comparison of \dedx{} Estimators}
\label{sec:tb:estimators}

Various methods exist to calculate the average \dedx{} of a track, that treat the Landau\-/tail differently.
The most commonly used method is the truncated mean, where a specified fraction of the hits with the highest \dedx{} is removed from each track for the calculation of the mean.
In some cases this can be improved by additionally removing a small fraction of the hits with the lowest \dedx{}.
This method is called trimmed mean.
Another method, adapted from~\cite{kleinwort2008}, relies on the transformation of the hit charges using \cref{eqn:tb:dedx_trans} to suppress the influence of the tail.
Increased computing power allows to try and fit the \dedx{} spectrum for each track.
Here, the challenge is to find the correct description of the distribution.
The Landau distribution, which is only valid for very short sample lengths and therefore does not describe the distribution in \cref{fig:hitcharge} well, was tested for comparison purposes.
A better fit is achieved by a log\-/normal distribution, but this does not match the tail.

\begin{figure}[htp]
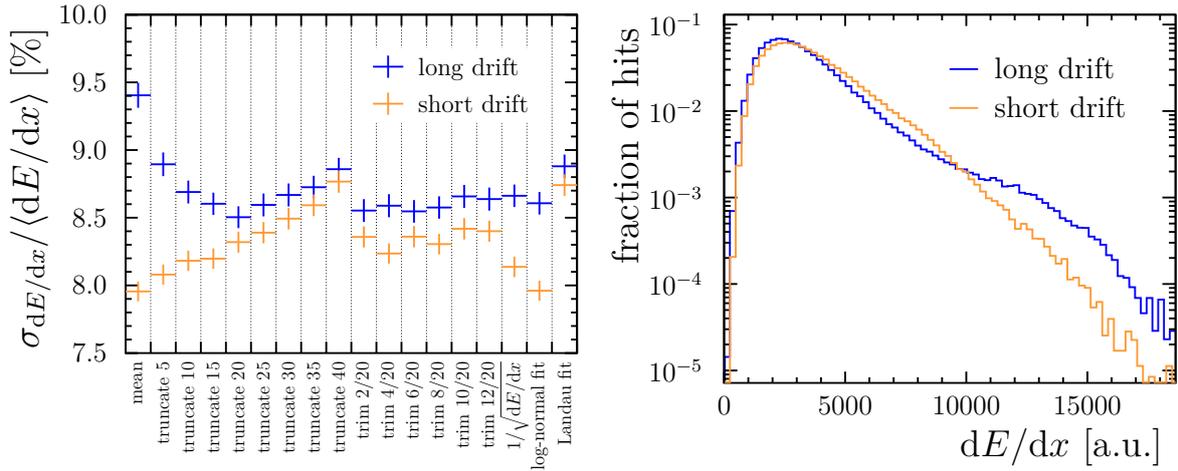

  \begin{subfigure}[t]{0.5\textwidth}
%     \centering
    \includestandalone[%
      width=\textwidth,
      height=0.33\textheight,
      keepaspectratio=true,
      ] {dEdx_est_cmp}
    \caption{Comparison of \dedx{} estimators.
      \label{fig:dedx_est_cmp}}
  \end{subfigure}%
  \begin{subfigure}[t]{0.5\textwidth}
    \includestandalone[%
      width=\textwidth,
      height=0.33\textheight,
      keepaspectratio=true,
      ] {dEdx_long_short_log}
    \caption{Hit \dedx{} distribution.
      \label{fig:dedx_dist_log}}
  \end{subfigure}
  \caption{\dedx{} at two drift distances (as given for \cref{fig:hit_cuts}).
    (\subref*{fig:dedx_est_cmp})~Performance of the different methods to calculate \dedx{} for tracks.
    For \emph{truncate~$X$} and \emph{trim~$Y/X$} the numbers $X$ and $Y$ indicate the cut\-/off fraction in percent at the upper and lower tail, respectively.
    (\subref*{fig:dedx_dist_log})~The difference in the resolution achieved with the same methods at the two drift distances is mainly due to the different shape of the \dedx{} distribution.
  }
%     \label{fig:}}
\end{figure}

A comparison of all methods for different cut\-/off fractions for the truncated and trimmed mean are shown in \cref{fig:dedx_est_cmp}.
The first thing to note is the large disparity between the two drift distances in the performance of the truncated mean at low truncation fractions.
As mentioned in \cref{sec:tb:analysis}, this is caused by the cut on saturated pulses rejecting less hits at higher drift distances due to the  larger spread of the charge over more pads.
In \cref{fig:dedx_dist_log}, the effect of this difference on the shape of the hit \dedx{} distribution can be seen.
For shorter drift distances below about \SI{100}{\milli\meter}, the tail is completely suppressed and the distribution is perfectly log\-/normal.
This leads to a much better better resolution from the low truncation fractions, the inverse\-/root transformation method and of course the log\-/normal fit.
Since this cut depends on the probability to find a saturated pulse in a hit, it also introduces a bias depending on the primary ionisation density and thus the particle species.
Therefore, rather than applying this cut, the fraction of saturated pulses should be kept low by reducing the amplification to achieve the best possible separation power.

For the long drift data with less saturated pulses, the best resolution of \SI{8.5+-0.1}{\percent} is achieved at a truncation cut\-/off fraction around \SI{20}{\percent}.
Additional trimming does not improve upon that.
From the two fitting functions, as expected, the log\-/normal performs better giving a resolution of \SI{8.6+-0.1}{\percent} compared to the Landau fit with \SI{8.9+-0.1}{\percent}.
Looking at the log\-/normal fit in the short drift data, where the tail is absent, shows what is possible with a function that matches the underlying distribution almost perfectly with a resolution of \SI{8.0+-0.1}{\percent}.
Generally, the fitting still increases the computing time by a factor of \num{10}, which may not be feasible for the scale of the full ILD data.
The inverse\-/root method with \SI{8.7+-0.1}{\percent} gives a slightly worse resolution than the best truncated mean.
Still it was decided to use this method, as it results in a more symmetric and Gaussian like \dedx{} distribution on hit level and therefore should be less biased.

\subsubsection{Extrapolation to the ILD TPC}
\label{sec:tb:extrapolation}

To estimate the performance of a full scale ILD TPC, the results from the prototype measurements need to be extrapolated.
Since the tail of the underlying energy loss distribution is cut off due to the finite dynamic range of the electronics, a scaling of $\sigma_{\mdedx} \propto 1/\sqrt{N_{hits}}$ with the number of hits $N_{hits}$ cannot be assumed.
Instead, a power law with an exponent $k<0.5$ is expected (see also \cref{eq_G013}):
\begin{equation}
  \label{eqn:dedx_res_scaling}
  \sigma_{\mdedx} \propto (N_{hits})^{-k}.
\end{equation}
Since the number of valid hits in the prototype is limited to \num{60} and does not vary greatly (see \cref{fig:num_hits}), an extrapolation to the \num{220} hits of the ILD TPC using only the existing tracks includes large uncertainties.
To circumvent this restriction, randomly selected hits from multiple tracks were combined to form pseudo\-/tracks of arbitrary length.
Since Marlin processes events sequentially, this is limited to hits from consecutive events.

\begin{figure}[!bp]
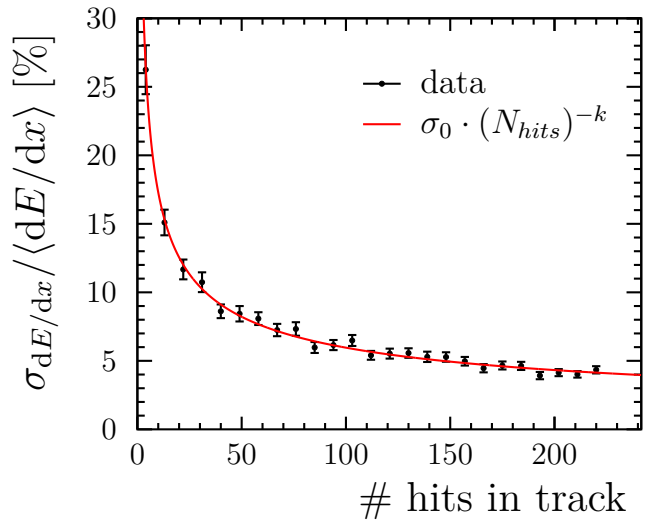

  \centering
  \includestandalone[%
    width=0.6\textwidth,
    height=0.33\textheight,
    keepaspectratio=true,
    ]{dEdx_res_vs_hits}
    \caption{The \dedx{} resolution achieved with pseudo\-/tracks of various lengths.
      The fitted power law has $\chi^2/ndf=\num{25.41/23}$.
      The values of the fit parameters are given in the text.
      \label{fig:extrapolation_data}}
\end{figure}

Applying this method the results shown in \cref{fig:extrapolation_data} were generated.
The power\-/law fit returns an exponent $k=\num{0.47+-0.01}$ and an intrinsic fluctuation for single hits $\sigma_0=\SI{53+-3}{\percent}$.
This leads to an estimate for the resolution at \num{220} hits of $\sigma_{\mdedx}(220)=\SI{4.2+-0.1}{\percent}$.
For the smaller version of the ILD TPC with only \num{165} rows the resolution is $\sigma_{\mdedx}(165)=\SI{4.8+-0.1}{\percent}$.
These values assume all hits are valid, i.e.\ from perfectly isolated and reconstructed tracks.
Taking into account about \SI{10}{\percent} invalid hits, both results increase by about \num{0.2} percentage points.

\section{High Granularity Readout Simulation}
In this section, the effect of a higher granularity of readout of a TPC on the \dedx{} resolution is investigated in simulation and quantified.

\subsection{Motivation}
 \label{sec_motivation}
As laid out in~\autocite[][chapter 10]{BlumRolandi08}, previous gaseous detectors have shown empirically that the achievable resolution of the energy loss measurement behaves like
\begin{equation}
  \sigma_{\mdedx} \propto L^{\num{-0.34}} \cdot N^{\num{-0.13}},
\end{equation}
with $L$ being the total length of a track in the TPC volume, and $N$ being the number hits of that track.
In common cases of comparison, e.g.\ different TPC sizes, both $L$ and $N$ are affected at the same time. Therefore it can be helpful to express the relation using the parameters pad height $H = L/N$ and granularity $G = 1/H = N/L$, leading to
\begin{equation}
  \sigma_{\mdedx} \propto L^{\num{-0.47}} \cdot G^{\num{-0.13}} = H^{\num{-0.34}} \cdot N^{\num{-0.47}}.
\label{eq_G013}
\end{equation}
Given a fixed granularity (i.e. pad height for a row\-/based readout), the dependence has the well know not\-/quite\-/square\-/root behaviour with an exponent of about \num{-0.47}.
This section, however, investigates the situation of a fixed track length $L$ and varying granularity, for which a weaker dependence with an exponent around \num{-0.13} is stated.
Since this is an empirical observation, comparing different detectors, it is not an exact theoretical value but rather gives the expected order of a dependence on the granularity.

%\subsection{Motivation for a highly granular hybrid readout}
\label{sec:idea}

\subsubsection{Cluster counting}

\begin{figure}[tbp]
  \begin{subfigure}{.5\textwidth}
    \centering
    \includegraphics[width=\textwidth,height=0.33\textheight,keepaspectratio=true]{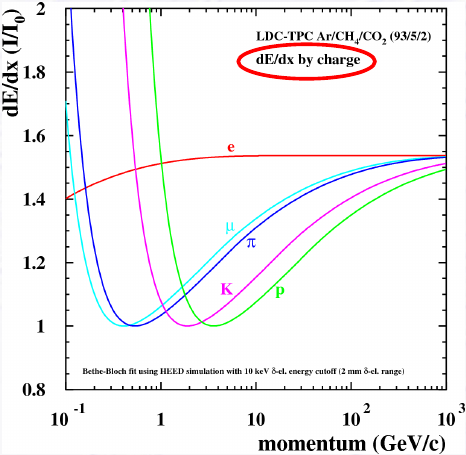}
    \caption{Bethe\-/Bloch curves.
      \label{fig:piokaonsep_BB}}
  \end{subfigure}%
  \begin{subfigure}{.5\textwidth}
    \centering
    \includegraphics[width=\textwidth,height=0.33\textheight,keepaspectratio=true]{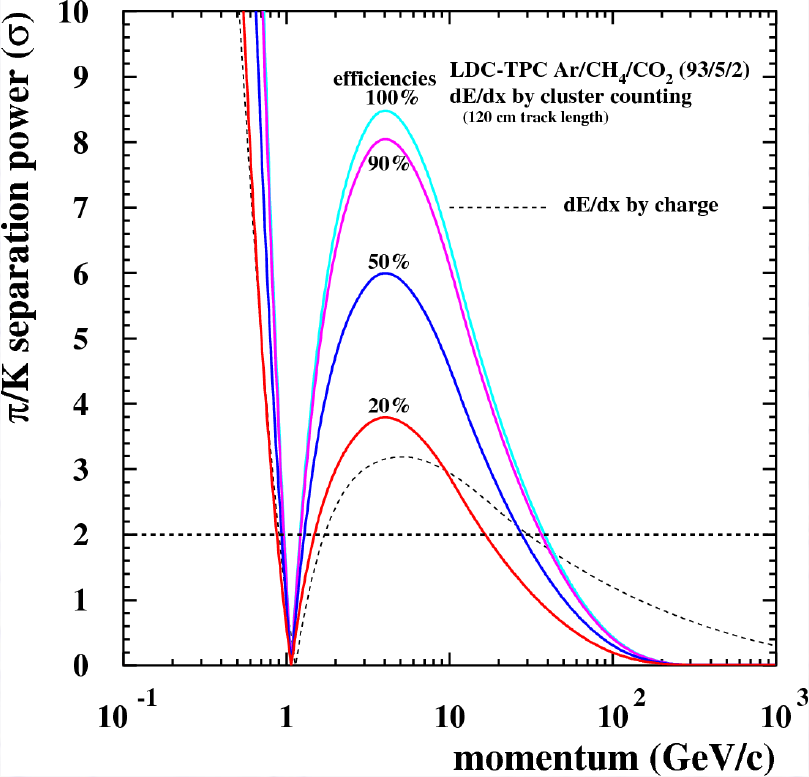}
    \caption{Pion/kaon separation power.
      \label{fig:piokaonsep_sep}}
  \end{subfigure}
  \caption{Theoretical expectation for Bethe\-/Bloch curves and pion/kaon separation power of \dedx{} by cluster counting, from \cite{Hauschild06}.}
  \label{fig:piokaonsep}
\end{figure}

The energy loss is conventionally determined from the deposited charge by summing all electrons generated from the ionisation by the incident particle.
The number of generated electrons in ionising interactions is given by a Landau distribution which has a long tail towards large values.
The relatively large width of this distribution worsens the correlation of the measured energy and the momentum of the particle.
It is advantageous to instead count the number of ionising interactions of the incident particle.
This follows a Poissonian distribution with a significantly smaller width, resulting in a better correlation and particle identification power, as shown in~\cite{Hauschild06}.
In \cref{fig:piokaonsep}, the separation power for pion/kaon\-/separation depending on the cluster counting efficiency is shown compared to the conventional \dedx{} by charge summation. In former experiments with prototypes, a cluster counting efficiency of only \SIrange{20}{30}{\percent} was reached. Nevertheless, the resulting separation power is still expected to be better than by using charge summation. Improved algorithms are expected to deliver a higher cluster counting efficiency. However, cluster counting can only work if a sufficient spatial correlation between the position of the electrons of one cluster is preserved during drift.

To achieve the cluster counting capability, a sufficiently high granularity is needed for the TPC readout system.
A possible hardware setup to achieve this, but at the same time keep the flexibility and large anode coverage of a pad-based readout is the ROPPERI system~\cite{ROPPERI17}.

\subsection{Software}
In this section, MarlinTPC is used to simulate a highly granular readout of the TPC drift volume, while part of the reconstruction is done with the source extractor software described below.
This makes it necessary to export the data to a different data format (.fits) and then re\=/import it to .slcio.
The processor chain consists of
\begin{enumerate}
  \item primary ionisation,
  \item drift,
  \item amplification with GEMs,
  \item digitisation with the Timepix ASIC,
  \item export to .fits,
  \item running of the source extractor software to reconstruct hits,
  \item import to .slcio,
  \item analysis (\dedx{} measurements, tracking).
\end{enumerate}

 \begin{figure}[htp]
   \centering
   \includegraphics[width=0.5\textwidth,height=0.33\textheight,keepaspectratio=true]{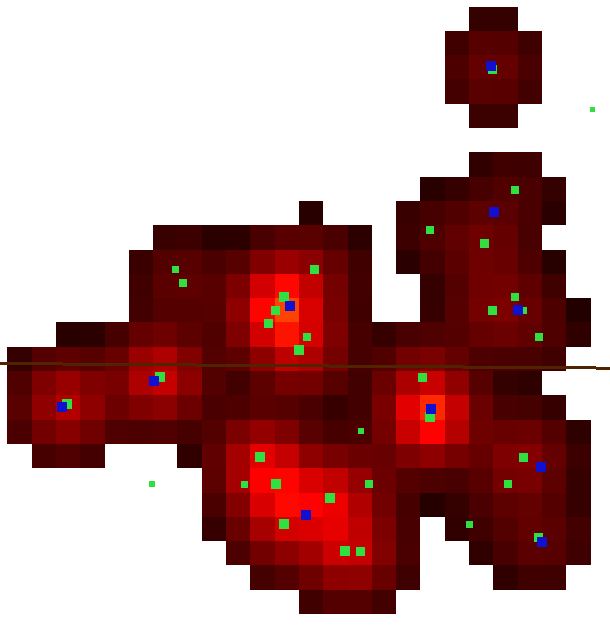}
   \caption{Example stages of the TPC\-/Timepix simulation and reconstruction:
    The brown line shows the track of the original MC particle.
    The green dots are the positions of the electrons from ionisation after drift (right before the first GEM).
    The red tiles show the level of activation of the underlying pixel chip from the charge clouds created by the GEM.
    The blue dots are the centre positions of the ``extracted sources'' based on the pixel charge information. They are in the following used as hits.
    \label{fig:sextract_exmpl}}
 \end{figure}

\subsubsection{Source Extractor}

Since the simulation targets a high granularity with pad sizes smaller than the charge clouds generated by the GEM amplification, the resulting 2\-/dimensional patterns of the readout are similar to astrophysical sky maps. There, often faint light sources in front of background light are investigated. One software to extract these sources from the sky maps is the source extractor software \cite{SourceExtractor}. It was first utilised and integrated into MarlinTPC by \cite{Deisting12}. It takes a 2\-/dimensional image, equivalent to a 2D\-/histogram, and computes the most likely position and brightness of sources, based on a provided typical shape.

The software uses a multi\-/staged approach, iteratively separating the image in sub\-/sections and adding sources to the fit. The sources are used as hits in the following steps, and are interpreted as clusters, albeit with a finite cluster identification and counting efficiency.
An example is shown in \cref{fig:sextract_exmpl}.

\subsection{Results}

\begin{figure}[tbp]
  \begin{subfigure}{.5\textwidth}
    \centering
    \includegraphics[width=\textwidth,keepaspectratio=true]{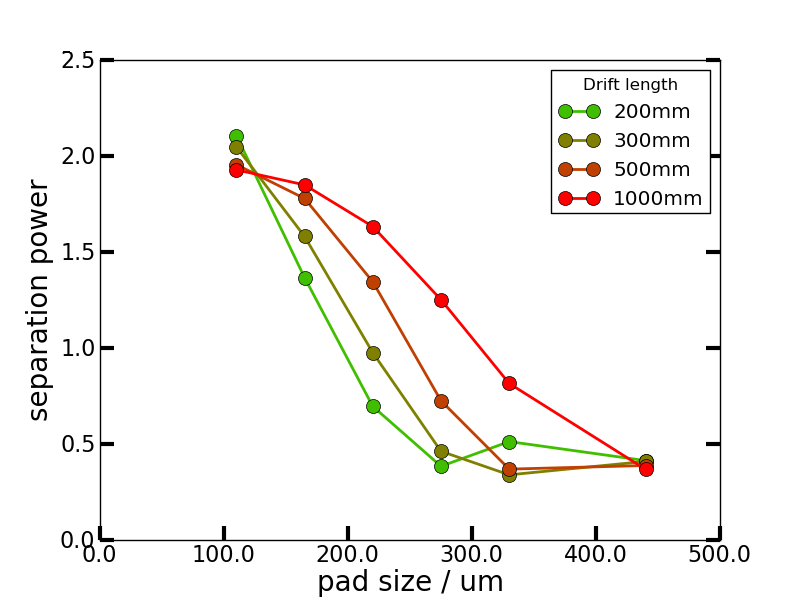}
    \caption{Fixed energy.
      \label{fig:res_seppow_PP_padsize}}
  \end{subfigure}%
  \begin{subfigure}{.5\textwidth}
    \centering
    \includegraphics[width=\textwidth,keepaspectratio=true]{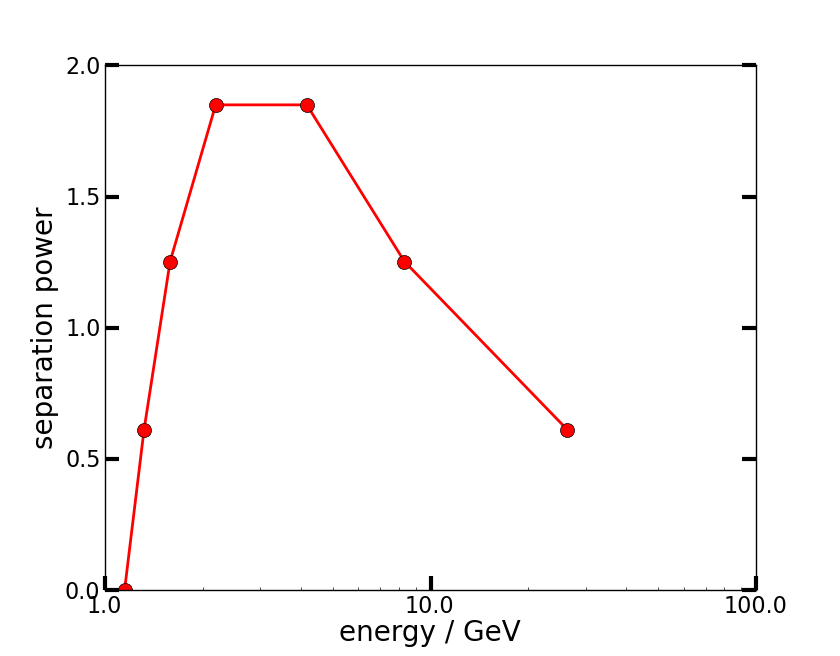}
    \caption{Fixed pad size and drift length.
      \label{fig:res_seppow_PP_momentum}}
  \end{subfigure}
  \caption{Simulation of separation power by cluster counting between pions and kaons
    (\subref*{fig:res_seppow_PP_padsize})~at the ionisation maximum of \SI{15}{\percent} depending on pad size and drift length and
    (\subref*{fig:res_seppow_PP_momentum})~depending on energy (and thus ionisation difference), with a pad size of \SI{165}{\micro\meter} and at a drift length of \SI{1000}{\milli\meter}.}
  \label{fig:res_seppow_PP}
 \end{figure}

Different TPC readout set ups were simulated, in particular varying the anode granularity.
The number of reconstructed clusters was counted and compared between a pion and a kaon, which have a local ionisation difference maximum of about \SI{15}{\percent} at a momentum around \SI{3}{\giga\electronvolt}.
Through saturation effects the number of reconstructed clusters is not exactly proportional to the amount of ionisation, so we use the separation power $S$ as a quantitative measure, defined as
\begin{equation}
  S = \frac{\abs{\mu_\pi-\mu_K}}{\sqrt{(\sigma_\pi^2+\sigma_K^2)/2}}.
\end{equation}
For a track length of \SI{300}{\milli\meter} and a magnetic field of \SI{1}{\tesla}, a separation power of around 2 could be achieved for very small pad sizes of \SI{110}{\micro\meter} as shown in \cref{fig:res_seppow_PP_padsize}.
Conversely, one can look at the separation power depending on the momentum of the pion and kaon, which is directly related to their ionisation difference.
The detailed simulation result in \cref{fig:res_seppow_PP_momentum} with a fixed pad size of \SI{165}{\micro\meter} and at a drift length of \SI{1000}{\milli\meter} reflects the earlier expectations in~\cite{Hauschild06} shown in \autoref{fig:piokaonsep}.

 \begin{figure}[htbp]
  \begin{subfigure}{.5\textwidth}
    \centering
    \includegraphics[width=\textwidth,keepaspectratio=true]{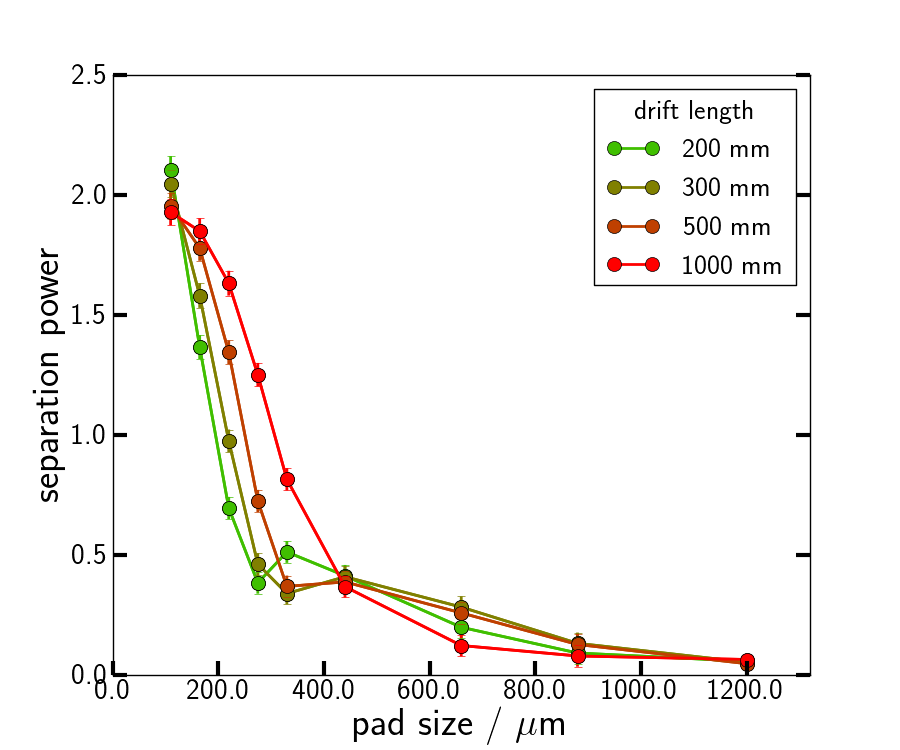}
    \caption{Cluster counting.
      \label{fig:res_seppow_methods_CC}}
  \end{subfigure}%
  \begin{subfigure}{.5\textwidth}
    \centering
    \includegraphics[width=\textwidth,keepaspectratio=true]{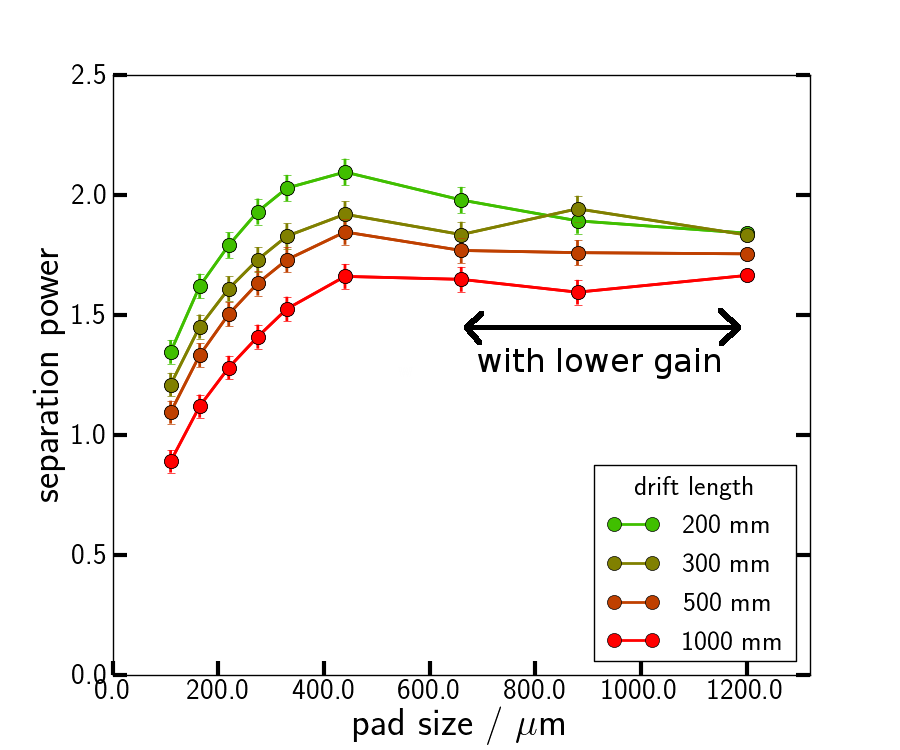}
    \caption{Charge summation.
      \label{fig:res_seppow_methods_charge}}
  \end{subfigure}
  \caption{Simulation of separation power between pions and kaons depending on pad size.
    \label{fig:res_seppow_methods}}
 \end{figure}

The \SI{300}{\milli\meter} tracks can be extrapolated to an ILD-like TPC with a track length of \SI{1.35}{\meter}, resulting in a separation power of $s=\num{3.8}$ for \SI{165}{\micro\meter} pads and $S=\num{3.4}$ for \SI{220}{\micro\meter} pads.
For this, combining several tracks gave a result compatible with a simple ($1/\sqrt{N}$)\-/extrapolation.
This result lies on the \SI{20}{\percent} cluster\-/counting\-/efficiency line in \cref{fig:piokaonsep_sep}, which is an improvement over the conventional charged based measure.
However, to achieve this, a very small pad size is necessary.
Extending the simulation parameter space to larger pads in \cref{fig:res_seppow_methods_CC} illustrates the drop at \SI{300}{\micro\meter}. Notably, in the transition region a larger drift distance is beneficial for the cluster counting, since at small drift distances a significant overlap between clusters masks many small ones, which can only be observed with very small pads.
The result for cluster counting can also be compared to the conventional charge summation, which is also expected to slowly improve for smaller pads with $\sigma_{\mdedx} \propto G^{\num{-0.13}}$, see \cref{eq_G013}.
In \cref{fig:res_seppow_methods_charge} this dependence of the separation power by charge summation can be observed, as long as the gain is adjusted to not run into electronics saturation for the larger pad sizes. For a fixed gain with very small pads threshold effects become relevant and limit the separation power, which is partly compensated at smaller drift lengths because the electrons are more concentrated.

 \begin{figure}[htp]
  \begin{subfigure}{.5\textwidth}
    \centering
    \includegraphics[width=\textwidth,keepaspectratio=true]{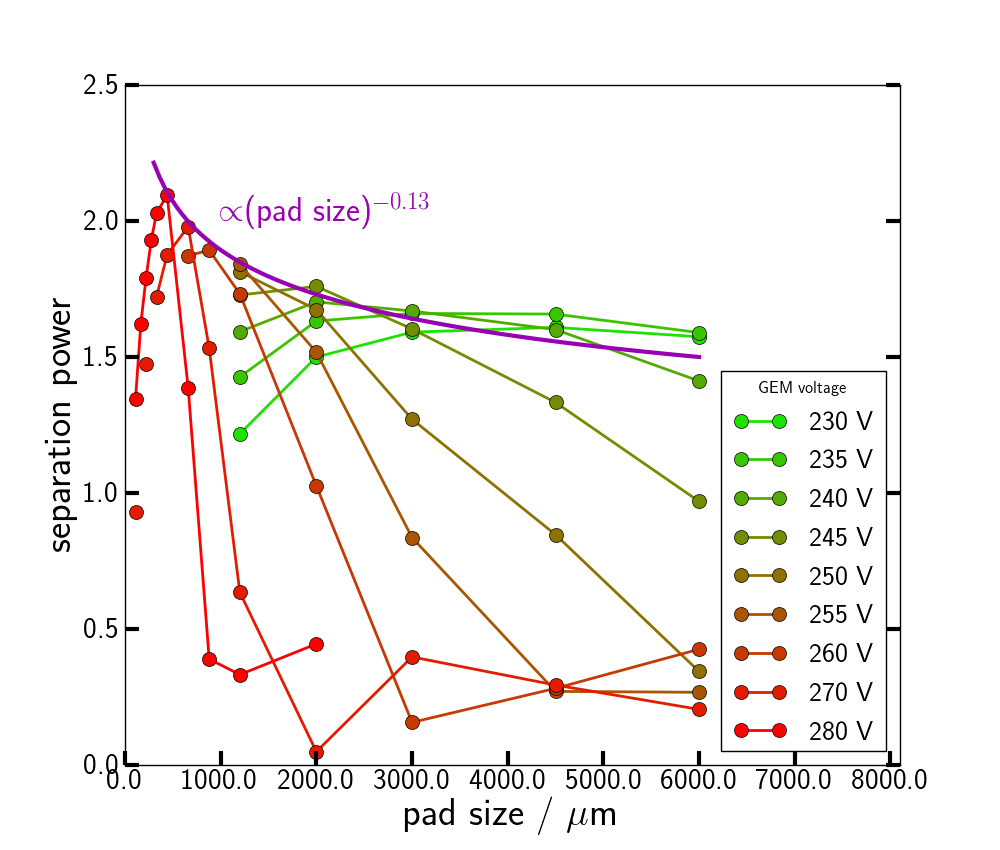}
    \caption{Optimising the GEM voltage.
      \label{fig:res_seppow_GUScan}}
  \end{subfigure}%
  \begin{subfigure}{.5\textwidth}
    \centering
    \includegraphics[width=\textwidth,keepaspectratio=true]{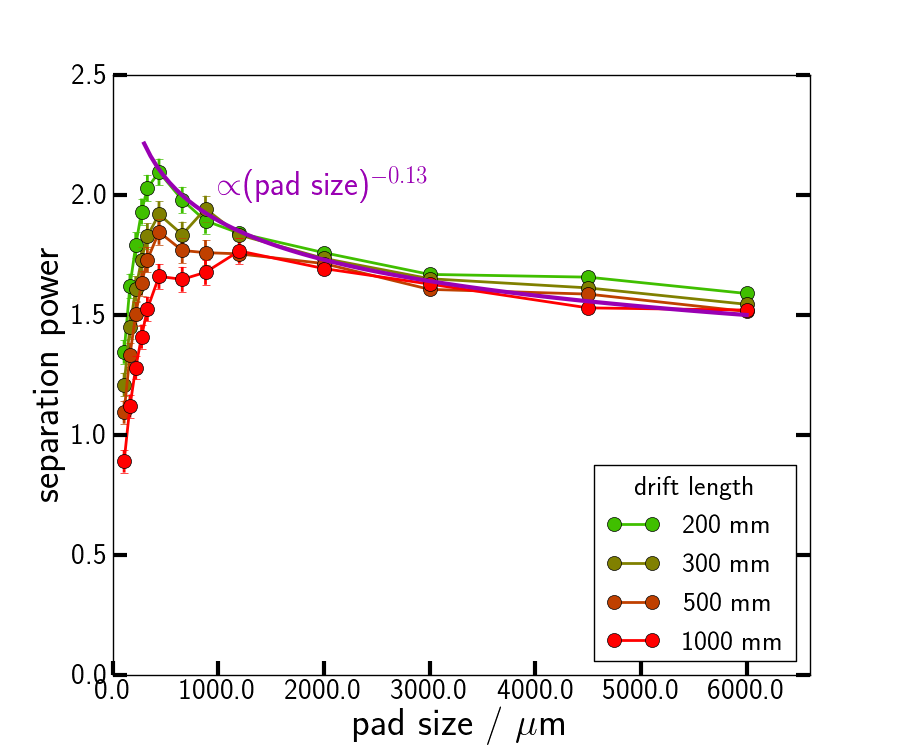}
    \caption{Result with optimal GEM voltage.
      \label{fig:res_seppow_GUScan_Res}}
  \end{subfigure}
  \caption{Simulation of separation power by charge summation between pions and kaons depending on pad size.
    (\subref*{fig:res_seppow_GUScan})~Scan of the GEM voltage, i.e.\ amplification gain, to find the optimum for each pad size.
    (\subref*{fig:res_seppow_GUScan_Res})~The result with optimised GEM voltage.
    \label{fig:res_seppow_GUScan_comb}}
 \end{figure}

%  \begin{figure}[htp]
%   \begin{subfigure}{.5\textwidth}
%     \centering
%     \includegraphics[width=\textwidth,keepaspectratio=true]{SP_GU_Scan_DL200}
%     \caption{\SI{200}{\milli\meter} drift length.
%       \label{fig:res_seppow_GUScan_200}}
%   \end{subfigure}%
%   \begin{subfigure}{.5\textwidth}
%     \centering
%     \includegraphics[width=\textwidth,keepaspectratio=true]{SP_GU_Scan_DL1000}
%     \caption{\SI{1000}{\milli\meter} drift length.
%       \label{fig:res_seppow_GUScan_1000}}
%   \end{subfigure}
%   \caption{Simulation of separation power by charge summation between pions and kaons depending on pad size and GEM voltage, i.e.\ gas amplification.
%     \label{fig:res_seppow_GUScan}}
%  \end{figure}

%  \begin{figure}[htp]
%   \begin{subfigure}{.5\textwidth}
%     \centering
%     \includegraphics[width=\textwidth,keepaspectratio=true]{SP_GUmax}
%     \caption{Linear x\=/axis.
%       \label{fig:res_seppow_GUScan_Res_lin}}
%   \end{subfigure}%
%   \begin{subfigure}{.5\textwidth}
%     \centering
%     \includegraphics[width=\textwidth,keepaspectratio=true]{SP_GUmax_log}
%     \caption{Logarithmic x\=/axis.
%       \label{fig:res_seppow_GUScan_Res_log}}
%   \end{subfigure}
%   \caption{Simulation of separation power by charge summation between pions and kaons with optimised GEM voltage, i.e.\ gas amplification.
%     \label{fig:res_seppow_GUScan_Res}}
%  \end{figure}

For the aforementioned reasons, at different pad sizes different gains were found to be optimal in terms of resulting separation power.
This was investigated for various GEM voltages in \cref{fig:res_seppow_GUScan} and lead to the results in \cref{fig:res_seppow_GUScan_Res}, if the optimal GEM voltage is chosen for each pad size.
A maximum gain of \num[tight-spacing=true]{64e3} at \SI{280}{\volt} was set for the simulation and it defines the cut\-/off at the smallest pad sizes.
Again, smaller drift distances allow for higher separation powers at small pad sizes because of threshold cut-off compensation.
In \cref{fig:res_seppow_GUScan_Res}, the empirical dependence on the pad size was added with an arbitrary scale factor and it matches with the simulation results well apart from the drop in the cut\-/off region.

 \begin{figure}[htp]
  \begin{subfigure}{.5\textwidth}
    \centering
    \includegraphics[width=\textwidth,keepaspectratio=true]{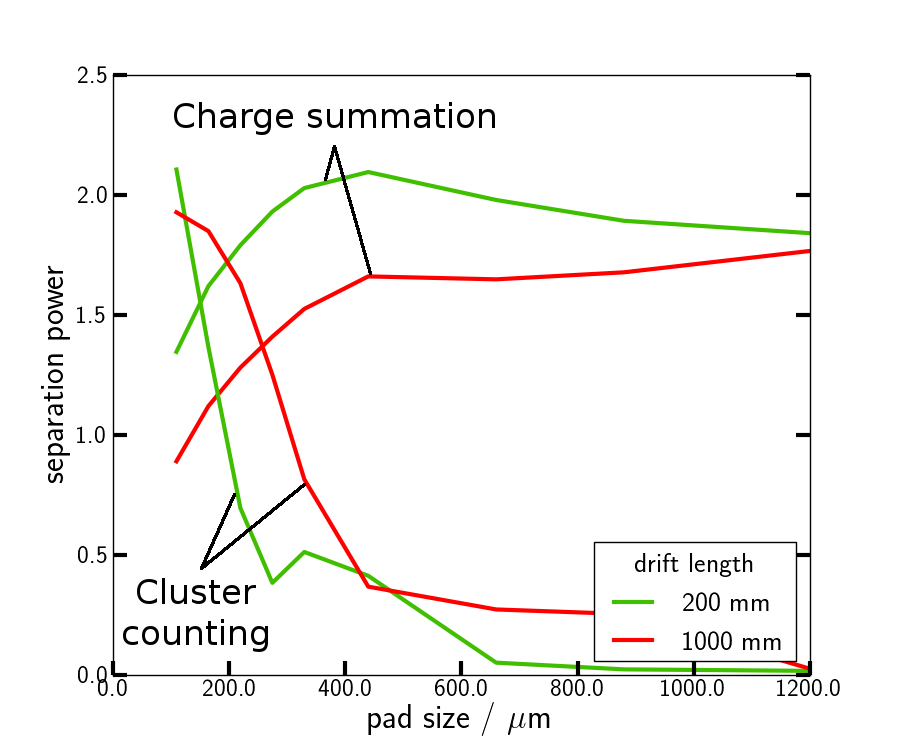}
    \caption{Linear x\=/axis.
      \label{fig:res_seppow_comb_lin}}
  \end{subfigure}%
  \begin{subfigure}{.5\textwidth}
    \centering
    \includegraphics[width=\textwidth,keepaspectratio=true]{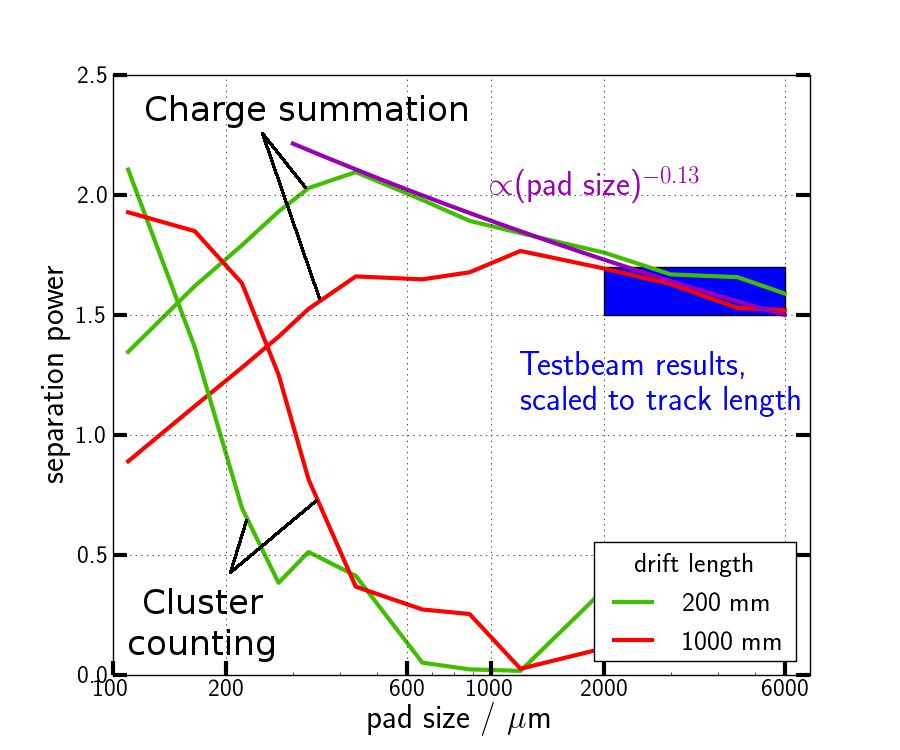}
    \caption{Logarithmic x\=/axis.
      \label{fig:res_seppow_comb_log}}
  \end{subfigure}
  \caption{Simulation of separation power between pions and kaons for cluster counting and charge summation.
    \label{fig:res_seppow_comb}}
 \end{figure}

Both, the separation power for cluster counting and the one for charge summation have been combined into the plots in \cref{fig:res_seppow_comb}. Only the smallest and largest simulated drift lengths are drawn.
To make them comparable to the existing systems, a blue band was inserted in \cref{fig:res_seppow_comb_log} to indicate the region that is covered by the beam test results, after transforming to separation power for \SI{300}{\milli\meter} long tracks, albeit for rectangular pads of \SI[product-units=brackets-power]{1x6}{\milli\meter} to \SI[product-units=brackets-power]{3x7}{\milli\meter}.
The separation power of the simulated charge summation is compatible with existing beam test results, and shows the expected rise to smaller pads.
Only at very small pad sizes below \SI{300}{\um} the cluster counting becomes effective, and for larger drift lengths above about \SI{500}{\mm} surpasses the maximum of the charge summation.

\section{Summary}
\label{sec:summary}

Using a large TPC prototype with about \num{53} valid hits on \SI{50}{\centi\meter} track length in a \SI{5}{\giga\electronvolt} electron beam, the \dedx{} resolution achieved with three GEM modules was measured to be \SI{8.7+-0.1}{\percent}.
Extrapolating this result to larger track lengths, the resolution in the large (220 rows) and small (165 rows) ILD TPC was estimated to be \SI{4.2+-0.1}{\percent} and \SI{4.8+-0.1}{\percent}, respectively, assuming \SI{100}{\percent} usable hits.
Taking into account \SI{10}{\percent} invalid hits, both values become only about \num{0.2} percentage points worse, staying at or below the stated goal for the ILD TPC of \SI{5}{\percent}.

Furthermore, it was shown that better \dedx{} resolution values can be achieved with a higher readout granularity.
The separation power between pions and kaons was calculated, using conventional charge summation, as well as using cluster counting.
Here, the amplification gain for a GEM\-/based system was optimised for the chosen pad size to compensate saturation and threshold cut\-/off effects.
The separation power peaks at a pad size around \SI{1x1}{\milli\meter}, limited by the realistically achievable amplification gain.
It follows a ($G^{\num{-0.13}}$)\-/dependence, which was empirically observed between different existing gaseous detectors.
At very small pad sizes below \SI{300}{\micro\meter} it becomes feasible to reconstruct GEM charge clouds using the source extractor software.
This enables to do cluster counting and has the potential to further increase the separation power for pad sizes below \SI{200}{\micro\meter} and larger drift lengths.

\section{Acknowledgements}

This material is based upon work supported by the National Science Foundation under Grant No.~0935316 and was supported by JSPS KAKENHI Grant No.~23000002. The research leading to these results has received funding from the European Commission under the 6th Framework Programme ``Structuring the European Research Area'', contract number RII3-026126, and under the FP7 Research Infrastructures project AIDA, grant agreement no.~262025.\\
Special thanks go to Y.~Makida, M.~Kawai, K.~Kasami and O.~Araoka of the KEK IPNS cryogenic group, and A.~Yamamoto of the KEK cryogenic centre for their support in the configuration and installation of the superconducting PCMAG solenoid.\\
The measurements leading to these results have been performed at the Test Beam Facility at DESY Hamburg (Germany), a member of the Helmholtz Association.\\
The authors would like to thank the technical team at the DESY II accelerator and test beam facility for the smooth operation of the test beam and the support during the test beam campaign. The contributions to the experiment by the University of Lund, KEK, Nikhef and CEA are gratefully acknowledged.

\printbibliography
\end{document}